\documentclass[12pt]{iopart}

\usepackage{graphicx}
\usepackage{iopams}

\newcommand \be{\begin{eqnarray}}
\newcommand \ee{\end{eqnarray}}
\newcommand \ba{\begin{eqnarray}}
\newcommand \ea{\end{eqnarray}}
\newcommand \bo{\boldsymbol}
\newcommand \nn{\nonumber}

\begin{document}

\title[The multiscale nature of streamers]
{The multiscale nature of streamers}

\author{U Ebert$^{1,2}$, C Montijn$^1$, T M P Briels$^2$, W Hundsdorfer$^1$,  
B Meulenbroek$^1$, A Rocco$^{1,3}$, E M van Veldhuizen$^2$} 

\address{$^1$ Centrum voor Wiskunde en Informatica (CWI), 
P O Box 94079, 1090GB Amsterdam, The Netherlands,\\
$^2$ Faculty of Physics, Eindhoven Univ. Techn., P O Box 513, 
5600MB Eindhoven, The Netherlands,\\
$^3$ University of Oxford, Dept. Statistics, 1 South Parks Road,
Oxford OX1 3TG, UK}

\ead{ebert@cwi.nl}

\begin{abstract}
Streamers are a generic mode of electric breakdown of large gas volumes. 
They play a role in the initial stages of sparks and lightning, 
in technical corona reactors and in high altitude sprite discharges
above thunderclouds. Streamers are characterized by a self-generated
field enhancement at the head of the growing discharge channel.
We briefly review recent streamer experiments and sprite observations.
Then we sketch our recent work on computations 
of growing and branching streamers, we discuss concepts and
solutions of analytical model reductions, we review different 
branching concepts and outline a hierarchy of model reductions.
\end{abstract}


\submitto{\PSST}

\maketitle

\section{Introduction}

Streamer discharges are a fundamental physical phenomenon 
with many technical applications 
\cite{EddieBook,Bogaerts,PitchW}. While their role in
lightning waits for further studies \cite{Raizer,Williams}, 
new lightning related phenomena above thunderclouds have been 
discovered in the past 15 years to which streamer concepts 
can be directly applied \cite{Sentman,Gerken,Pasko,PasNat}.

In past years, appropriate methods have become available 
to study and analyze these phenomena.
Methods include plasma diagnostic methods, large scale computations
as well as analysis of nonlinear fronts and moving boundaries.
The aim of the present article is to briefly summarize progress
in these different disciplines, to explain the mutual benefit
and to give a glimpse on future research questions.

The overall challenge in the field is to understand the growth of
single streamers as well as the conditions of branching or extinction
and their interactions which would allow us to predict the overall
multi-channel structure formed by a given power supply in a given gas.
We remark that so-called dielectric breakdown models \cite{DBM1,DBM2} 
have been suggested to address this question, but they incorporate
the underlying mechanisms on smaller scales in a too qualitative way.

Many of our arguments are
qualitative as well, but in a different sense. The main theoretical
interest of the present paper is in basic conservation laws, 
in the wide range of length and time scales that characterize 
a streamer, in physical mechanisms for instabilities, and 
in the question whether a given problem should be modelled 
in a continuous or a discrete manner. Answering these
questions provides the basis for future quantitative predictions.

The paper is organized as follows:
In Section 2 a short overview over current experimental questions,
methods and results is given, and applications are briefly discussed.
Sprite discharges above thunderclouds are reviewed. 
In Section 3, the microscopic mechanisms and modelling issues 
are discussed and characteristic scales are identified by dimensional
analysis. In Section 4 numerical
solutions for negative streamers in non-attaching gases are 
presented, the multiple scales of the process are discussed and a
short view on numerical adaptive grid refinement is given.
Section 5 summarizes an analytical model reduction
to a moving boundary problem, sketches issues of charge conservation
and transport and confronts two different concepts of streamer branching.
Conclusion and summary can be found in Section 6.


\section{Streamers and sprites, experiments, applications and observations}

The emergence and propagation of streamers has a long research
history. The basics of a theory of spark breakdown were developed
in the 1930'ies by Raether, Loeb and Meek 
\cite{Raether,LoebMeek}.

\subsection{Time resolved streamer measurements}

The first experiments on streamers were carried out by Raether who
took pictures of the development of a streamer in a cloud 
chamber~\cite{Raether}.
In this experiment a discharge was generated by a short voltage pulse.
The ions within the streamer region act as nuclei for water droplets
that form in the cloud chamber. Photographs of these droplets 
show the shape of the avalanche or the emerging streamer.

Later the use of streak 
photography together with image intensifiers enabled researchers 
to take time resolved pictures of streamers~\cite{wag1966,wag1967,cha1972}.
Streak photographs show the evolution of a slit-formed section
of the total picture as a function of time. Examples of streak 
photographies can be found in this volume \cite{Williams}.

\begin{figure}
\begin{center}
\includegraphics[height=12cm]{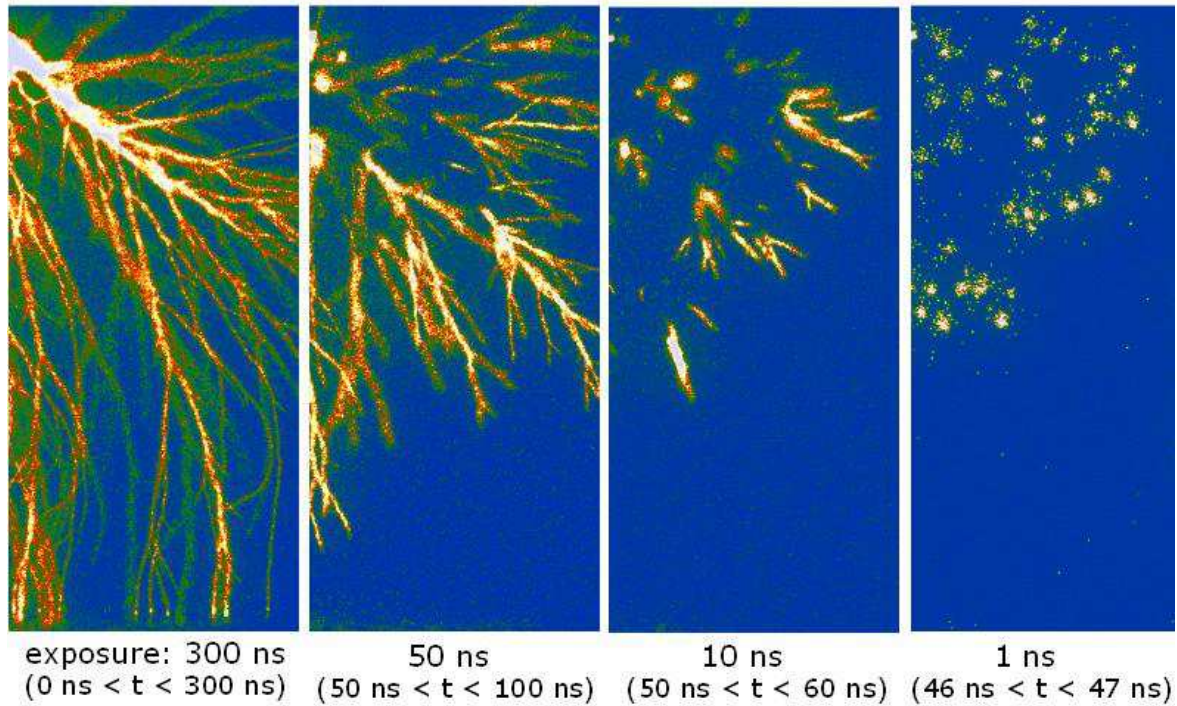}
\caption{ICCD photographs in the present Eindhoven experiments
of positive streamers in ambient air between point electrode 
in the upper left corner and plane electrode below. Only the right halves
of the figures are shown. The gap spacing is 4 cm, 
the applied voltage is 28 kV.
The exposure times of the photographs is 300, 50, 10 and $\sim$1 ns,
the actual time intervals are given in brackets where $t=0$ is the 
approximate time when the streamers emerge from the upper point.
}
\label{vel}
\end{center}
\end{figure}

Recently, ICCD (Intensified Charge Coupled Device) 
pictures yield very high temporal resolution, 
and at the same time a full picture of the
discharge, rather than the one-dimensional subsection obtained with
streak photography. First measurements since 1994 with 30 ns~\cite{Crey} and 
5 ns~\cite{Blom} resolution showed the principle. Since 2001, a resolution 
of about 1 ns has been 
reached~\cite{vel2001-2,vel2002,vel2002-2,vel2002-3,YiWilliams,IEEETanja}. 
Meanwhile, even shorter gate times are possible \cite{pan2005}.
However, the C-B transition of the second positive system of N$_2$ 
is the most intensive and dominates the picture, and its lifetime 
is of the order of 1 to 2 ns at atmospheric pressure. Therefore, 
a further improvement of the temporal resolution of the camera
does not improve the temporal resolution of the picture, 
but has to be payed with a lower spatial resolution and 
a lower photon number density.
Our measurements with resolution down to about 1 ns therefore resolve 
the short time structure of streamers at atmospheric pressure 
down to the physical limit. Fig.\,\ref{vel} shows snapshots of positive 
streamers in ambient air emerging from a positive point electrode 
at the upper left corner of the picture and extending to a plane 
electrode at the lower end of the picture. The distance between 
point and plane is 4 cm and the applied voltage about 28 kV.
The optics resolves all streamers within the 3D discharge, also within 
the depth. The filamentary structure of the streamers as well as their 
frequent branching is clearly seen on the left most picture 
with 300 ns exposure time. The rightmost picture has the shortest 
exposure time of 1 ns. It shows not the complete streamers, but only
the actively growing heads of the channels where field and impact
ionization rates are high. As a consequence, the other pictures 
have to be interpreted not as glowing channels, but as the trace 
of the streamer head within the exposure time. Streamer velocities 
can therefore directly be determined as trace length devided by
exposure time. 

Most experimental work has been carried out on positive streamers in
air~\cite{vel2002,IEEETanja,pan2005,bri2005-2}. We wish to draw the reader's
attention to the work by Yi and Williams~\cite{YiWilliams}, who
investigated the propagation of both anode and cathode-directed streamers, 
in almost pure N$_2$ and in N$_2$/O$_2$ mixtures. We will briefly interprete 
their results in Section 3.

\begin{figure}
\begin{center}
\includegraphics[height=6cm]{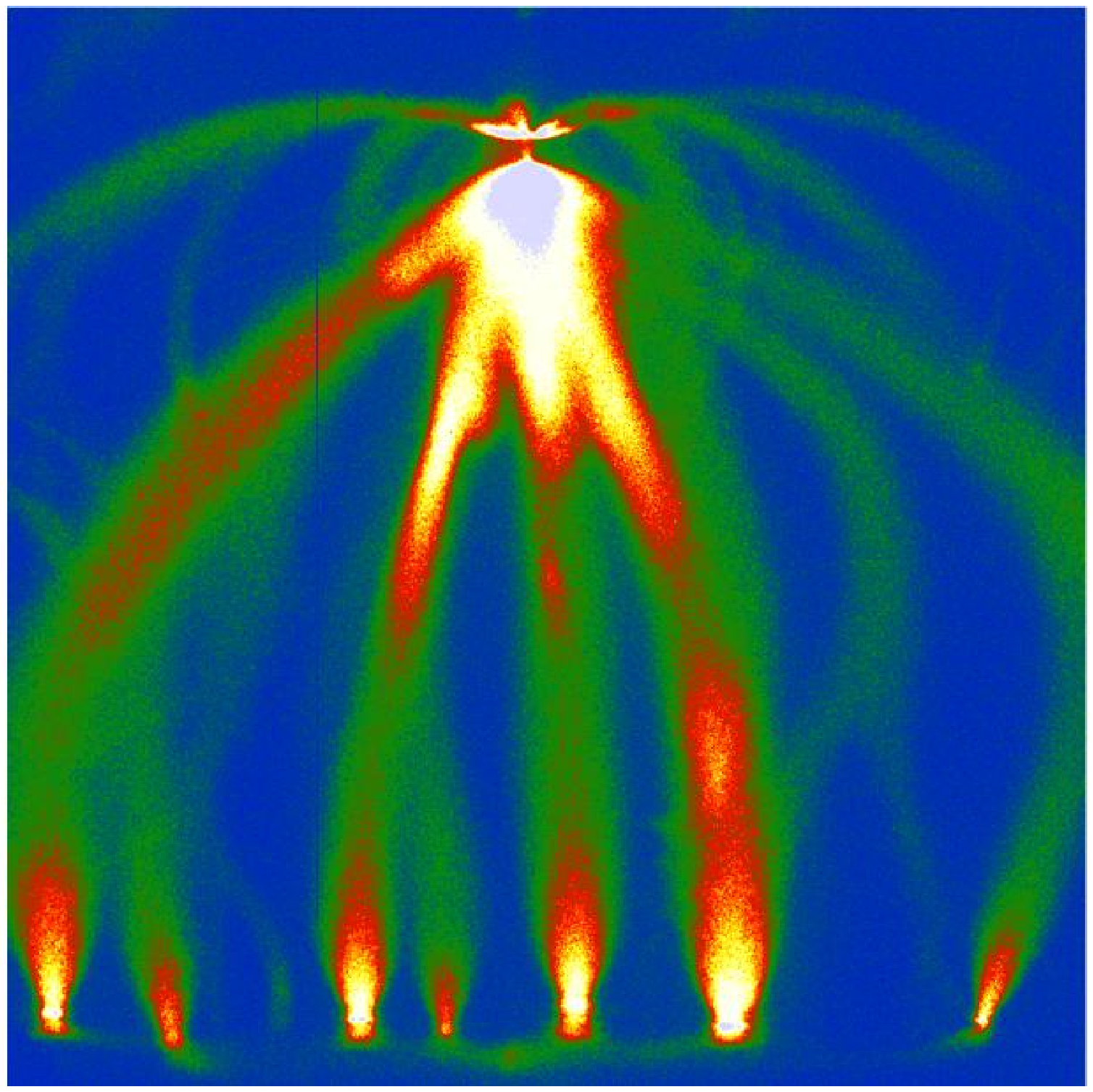}
\caption{ICCD photograph of a positive streamer in ambient air
in the same electrode configuration as in Fig.~1, but now powered
with a transmission line transformer and a peak voltage of 60 kV
\cite{Thick}.
The voltage pulse lasts about $\sim100$ ns and limits the duration
of the discharge; the exposure time of 20 $\mu$s accounts 
for the jitter of the voltage switching.
}
\label{thick}
\end{center}
\end{figure}

The experimental results depend not only on applied voltage,
but also on further features of the external power supply. 
A glimpse is given in Fig.~2. For a further discussion of this feature,
we refer to \cite{IEEETanja,pan2005,bri2005-2} and future analysis.
Furthermore the results depend on the gas pressure \cite{IEEETanja} 
as will be further discussed below.


\subsection{Applications}

There are numerous applications of streamers in corona discharges
\cite{EddieBook}. 
Dust precipitators use DC corona to charge small particles and 
draw them out of a gas stream. This process is used in industry 
for already more than a century. Another wide spread application 
is charging photoconductors in copiers and laser printers. 
The first use of a pulsed discharge has been the production of ozone 
with a barrier configuration in 1854. This method is still being used 
\cite{EddieBook} but pulsed corona discharges obtain the same ozone yield 
\cite{E25}. 

In the 1980's the chemical activity of pulsed corona was recognized 
and investigated for the combined removal of SO$_2$, NO$_x$ and fly ash 
\cite{E26,E27}. In the same period also the first experiments on water 
cleaning by pulsed corona were performed \cite{E28}. More results 
on combined SO$_2$/NO$_x$ removal can be found in \cite{EddieBook}, 
recent results on degradation of phenol in water are given 
in \cite{E25} and \cite{E29}. 
More recently, the chemical reactivity is further explored for example 
in odor removal \cite{E30}, tar removal from biogas \cite{E31} 
and killing of bacteria in water \cite{E32}. 

A new field is the combination 
of chemical and hydrodynamic effects. This can be used in 
plasma-assisted combustion \cite{E33} and flame control \cite{E34}. 
Purely electrohydrodynamic forces are studied in applications 
such as aerodynamic flow acceleration \cite{PitchW,StariW} for aviation
and plasma-assisted mixing \cite{E35}.

Basically, these applications are based on at least one of three principles:
1) the deposition of streamer charge in the medium, 2) molecular excitations
in the streamer head that initiate chemical processes, 
and 3) the coupling of moving space charge regions to gas convection
\cite{PitchW}. The chemical applications are based on the exotic 
properties of the plasma in the streamer head that acts as 
a self-organized reactor: a space charge wave carries a confined 
amount of high energetic electrons that effectively ionize
and excite the gas molecules. It is this active region that is seen
in ICCD pictures like Fig.~1.


\subsection{Sprite discharges above thunderclouds}

Streamers can also be observed in nature. They play a role 
in creating the paths of sparks and lightning~\cite{Raizer,Williams,maz1995}, 
and sensitive cameras showed the existence of so-called
sprites~\cite{Sentman,fra1990,lyo1996}
and blue jets~\cite{Pasko,wes1995,wes1998} in
the higher regions of the atmosphere above thunderclouds. 
With luck and experience, sprites can also be seen with naked eye.
A scheme of sprites, jets and 
elves as the most frequently observed lightning related
transious luminous events above lightning clouds is given in Fig.~3a. 

\begin{figure}[h]
\begin{center}
\includegraphics[height=6.5cm]{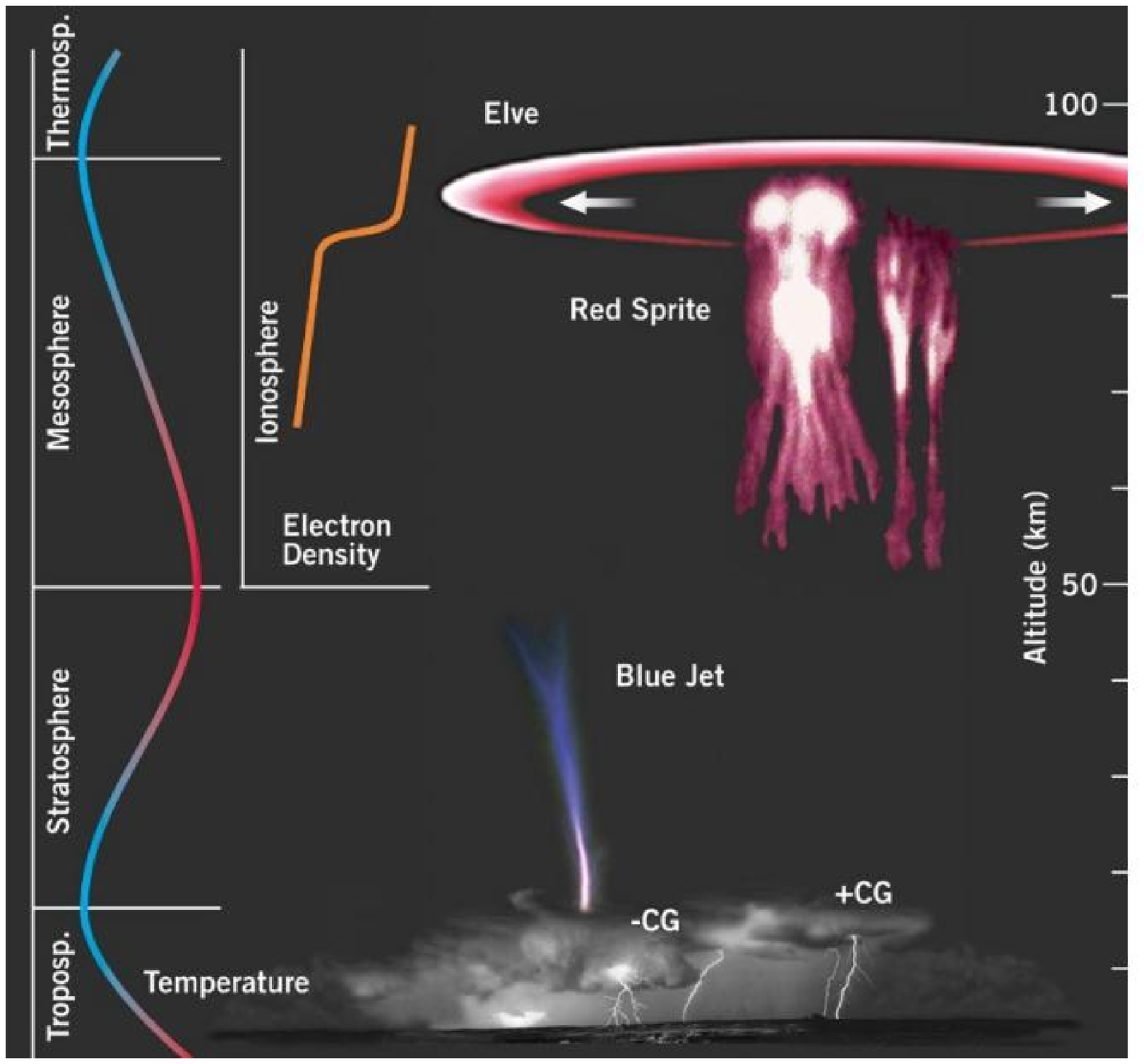}
\includegraphics[height=6.5cm]{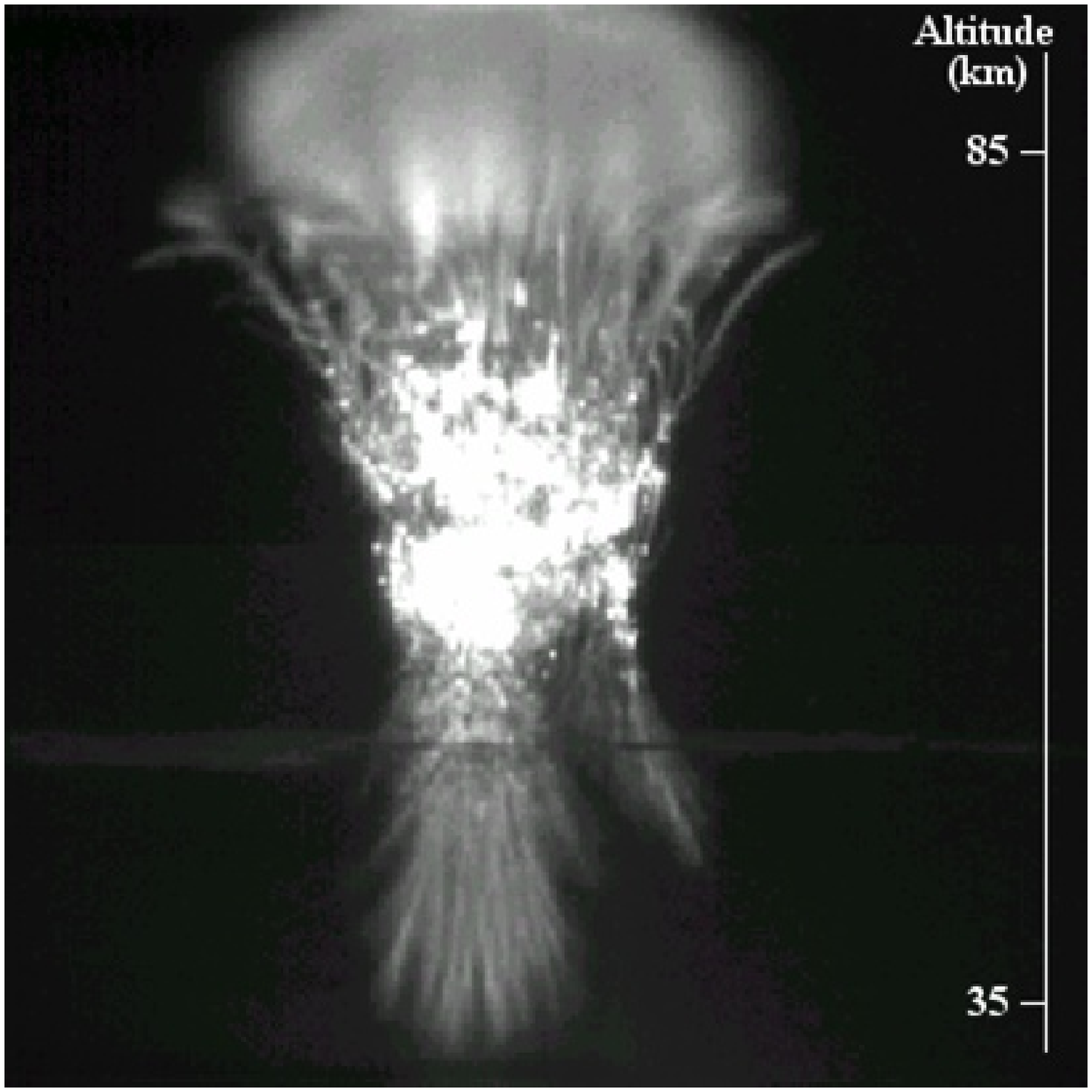}
\caption{a) Scheme of transient luminous events above thunderclouds
\cite{Neubert}. b) Photograph of a sprite. The altitude is indicated 
on the right. 
Courtesy of H.C. Stenbaek-Nielsen, Dept. Geophys., Univ. Fairbanks, Alaska.} 
\label{ste}
\end{center}
\end{figure}

\begin{figure}
\begin{center}
\includegraphics[width=13cm]{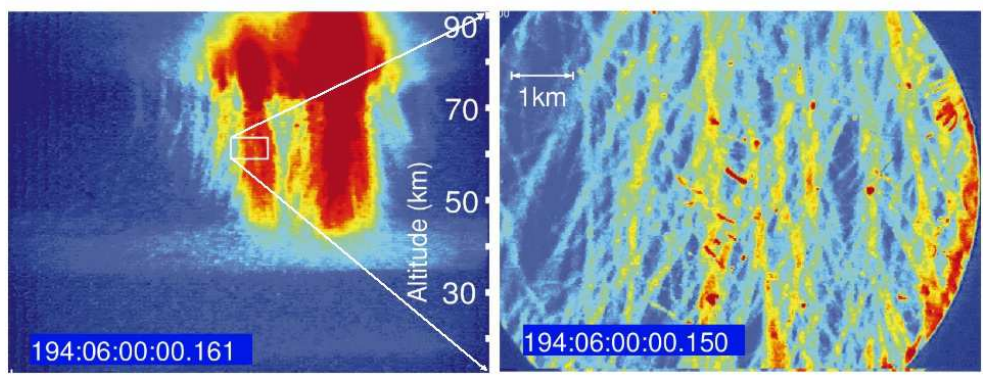}
\caption{Telescopic image of a sprite discharge. The right photograph
zooms into the white rectangle of the left photograph. These pictures 
are taken from~\cite{Gerken}.} 
\label{ger}
\end{center}
\end{figure}

Sprites have been observed between about 40 and 90 km height 
in the atmosphere. Above about 90 km, solar radiation maintains 
a plasma region, the so-called ionosphere. In this region,
so-called elves can occur which are expanding rings created
by electromagnetic resonances in the ionospheric plasma.
Sprites, on the other hand, require a lowly or non-ionized medium,
they can propagate from the ionosphere downwards or from
some lower base upwards, or they can emerge at some immediate
height and propagate upwards and downwards like in the event shown 
in Fig.~3b. A variety of sprite forms have been reported
\cite{Gerken,StenbaekNielsen}.
The propagation direction and approximate velocity 
-- their speed can exceed 10$^7$ m/s~\cite{Stanley} --,
is known since researchers succeeded in taking movies \cite{PasNat}.
A sequence of movie pictures can be found in this volume \cite{Williams}.
Telescopic images of sprites show that they are composed
of a multitude of streamers (see Fig.~\ref{ger}).
Blue jets propagate upwards from the top of thunderclouds, 
at speeds that  are typically two orders of magnitude lower 
than those of sprites, and they have a characteristic conical shape 
and  appear in a blueish color~\cite{wes1995,wes1998}.
Sprite discharges are the most frequent of these phenomena.

The approximate similarity relations between streamers and sprites 
are discussed in the next section. Sprites could therefore have similar
physical and chemical effects as those discussed for streamer
applications in the previous section. In fact, charge deposition in the
medium and molecular excitations with subsequent chemical processes
can be expected as well. On the other hand, we will show below 
by dimensional analysis that generation of gas convection 
is unlikely in sprites.
                                                             
\section{Microscopic modelling}

\subsection{An overview of modelling aspects}
                                                                                                    
The basic microscopic ingredients for a streamer discharge are\\
1) the generation of electrons and ions in regions of high electric field,\\
2) drift and diffusion of the electrons in the local field, and \\
3) the modification
of the externally applied field by the generated space charges.
\\
While avalanches evolve in a given background field,
streamers have a characteristic nonlinear coupling between densities
and fields: The space charges change the field, and the field
determines the drift and reaction rates.
More specifically, the streamer creates a self-consistent field
enhancement at its tip which allows it to penetrate into regions
where the background field is too low for an efficient ionization
reaction to take place. In this sense, streamers are similar
to mechanical fractures in solid media: either the electric 
or the mechanical forces focus at the tips of the extending structures.
                            
While these general features are the same, models vary
in the following aspects:\\
$\bullet$ the number of species and reactions included 
in a model~\cite{dha1987,Vitello,Babaeva,Pan2003},\\
$\bullet$ the choice for ``fluid models'' with continuous particle
densities~\cite{dha1987,Babaeva,PRLMan,PREStrAndr,Pan2003,IEEECaro} 
versus models that trace single 
particles or super-particles~\cite{qin1994,sor2001}\\
$\bullet$ local versus nonlocal modelling of drift and ionization rates
(by electron impact and photo-ionization), and \\
$\bullet$ assumptions about background ionization from natural
radioactivity or from previous streamer events in a pulsed streamer mode,
and choices of electrode configuration, plasma-electrode interaction,
initial ionization seed and external 
circuit~\cite{dha1987,Vitello,Babaeva,pan2001,Pan}.\\
$\bullet$ There are so-called 1.5-dimensional models that include
assumptions about radial properties into an effectively 1D numerical
calculations~\cite{dav1967,mor1997}, so-called 2D models, 
that solve the 3D problem assuming cylindrical 
symmetry~\cite{dha1987,Vitello,Babaeva,PRLMan,PREStrAndr,Pan2003,IEEECaro}, 
and a few results on fully 3D models have been reported~\cite{kul1998,Pan}.

It also should be recalled, that the result of a numerical computation
does not necessarily resemble the solution of the original equations:
results can depend on spatial grid spacing and time stepping,
on the computational scheme etc. Adding many more species with not
well-known reaction rates might force a computation to use a lower spatial
resolution and lead to worse results than a reduced model. Model
reduction techniques therefore should be applied to problems with a
complex spatio-temporal structure like streamers;
they can be based on a large difference of inherent length
and time scales, key techniques are adiabatic elimination, singular
perturbation theory etc.
Furthermore, considering a single streamer with cylindrical
symmetry, the results of a 2D calculation will be numerically
much more accurate than those of a 3D calculation, and a fluid
or continuum approximation should be sufficient. On the other hand,
a really quantitative model of streamer branching
should include single-particle statistics (not super-particles!)
in the leading edge of the 3D ionization front.
All in all, it can be concluded that the model choice also
depends on the physical questions to be addressed.

\subsection{The minimal model}

All quantitative numerical results in the present paper are
obtained for negative streamers with local impact ionization reaction
in a fluid approximation for three species densities: 
the electron density $n_e$ and the densities of positive and negative
ions $n_\pm$ coupled to an electric field $\bo{\cal E}$
in electrostatic aproximation ${\bo{\cal E}}=-\nabla_{\bf R}\Phi$.
The model reads
\be
\partial_\tau\,n_e&=&  D_e\nabla_{\bf R}^2n_e 
  + \nabla_{\bf R}\cdot\left(\mu_e\, n_e\,{\bo{\cal E}}\right) \nn\\ 
  & &+(\mu_e\,|{\bo{\cal E}}|\,\alpha(|{\bo{\cal E}}|)
       -\nu_a)\,n_e,\nn\\
\partial_\tau\,n_+&=&\mu_e\,{\bo{\cal E}}\,\alpha({\bo{\cal E}})\,n_e ,\nn\\
\partial_\tau\,n_-&=&\nu_a\, n_e,\nn\\
\nabla_{\bf R}^2\Phi&=&\frac{\rm e}{\epsilon_0}\;(n_e+n_--n_+)~~~,~~~
{\bo{\cal E}}=-\nabla_{\bf R}\Phi .
\label{model}
\ea
Here $D_e$ and $\nu_a$ are the electron diffusion coefficient 
and the electron attachment rate, $\mu_e$ and $D_e$ are the electron 
mobility and diffusion constant. We assume that the impact ionization rate 
$\alpha(\bo{\cal E})$ is a function of the electric field, and
that it defines characteristic scales for cross-section $\alpha_0$
and field strength ${\cal E}_0$, similarly as in the 
Townsend approximation $\alpha(|{\bo{\cal E}}|)=
\alpha_0\exp(-{\cal E}_0/{|\bo{\cal E}}|)$. (The Townsend approximation
together with $\nu_a=0$ is actually used in our presented computational
results.) 
Electrons drift in the field and diffuse, while positive and negative 
ions are considered to be immobile on the time scales investigated 
in this paper. 

\subsection{Dimensional analysis}

Dimensionless parameters and fields are introduced as 
\begin{eqnarray}
{\bf r}= \alpha_0\;{\bf R},&~~~& t=\alpha_0\mu_e{\cal E}_0\;\tau, \nn\\
\sigma=\frac{{\rm e}\;n_e}{\epsilon_0\alpha_0{\cal E}_0},&~~~ 
& \rho=\frac{{\rm e}\;(n_+-n_-)}{\epsilon_0\alpha_0{\cal E}_0}, \nn\\
{\bf E}=\frac{\bo{\cal E}}{{\cal E}_0},&~~~ &f(|{\bf E}|,\nu)
=|{\bf E}|\;\frac{\alpha(|{\bf E}|\;{\cal E}_0)}{\alpha_0}-\nu,\nn\\
D=\frac{D_e\alpha_0}{\mu_e{\cal E}_0},&~~~
& \nu=\frac{\nu_a}{\alpha_0\mu_e{\cal E}_0},
\end{eqnarray}
which brings the system of equations (\ref{model}) into the dimensionless form
\ba
\label{nodim}
\partial_t\,\sigma & = &
   D\nabla^2\sigma+ \nabla(\sigma{\bf E}) +
   f(|{\bf E}|,\nu)\,\sigma \label{sigmand}\, ,\\
\partial_t\,\rho & = & f(|{\bf E}|,\nu)\,\sigma \label{rhond}\, ,\\
\nabla^2\phi & = & -\nabla\cdot{\bf E}=\sigma-\rho\, .
\label{poissnd}
\ea
It is interesting to note that even for several charged species 
$n_e$ and $n_\pm$, the computations can be reduced to only two density fields:
one for the mobile electrons $\sigma$ and one for the space charge
density of all immobile ions $\rho$.

The intrinsic parameters depend on neutral particle density $N$ 
and temperature $T$, for N$_2$ they are with the parameter values 
from \cite{dha1987,Vitello} given by
\be
\label{Sc1}
\alpha_0^{-1}\approx\frac{2.3~{\rm \mu m}}{N/N_0},&&~~~
{\cal E}_0\approx200 ~\frac{\rm kV}{\rm cm}\;\frac{N}{N_0},
\\
\mu_e{\cal E}_0\approx 760 \;\frac{\rm km}{\rm s}=760 \;\frac{\mu \rm m}{\rm ns},
&&~~~
D\approx 0.12\;\frac{T}{T_0},
\\
\label{Sc2}
\frac{\epsilon_0\alpha_0{\cal E}_0}{\rm e}
\approx \frac{5\cdot 10^{14}}{{\rm cm}^3}\;\left(\frac{N}{N_0}\right)^2&=&
2\cdot10^{-5}\frac{N}{N_0}\;N,
\ee
where $N_0$ and $T_0$ are gas density and temperature under normal conditions.

Based on this dimensional analysis, it can be stated immediately
that the ionization within the streamer head will be of the order 
of $10^{14}$/cm$^{-3} (N/N_0)^2$, the velocity will be of the order
1000 km/s etc. The natural units for length, time, field and
particle density depend on gas density $N$, the dimensionless 
diffusion coefficient depends on temperature $T$ through the Einstein relation,
and the characteristic streamer velocity is independent of both $N$ and $T$.
As far as the minimal model is applicable, these equations identify 
the scaling relation between streamers and sprites. While lab streamers
propagate at about atmospheric pressure, sprites at a height of
70 km propagate through air with density about 5 orders of magnitude lower. 
Sprite streamers are therefore about 5 decades larger (one streamer-cm
corresponds to one sprite-km etc.), but their velocities are similar.

It should be noted that all particle interactions taken
into account within the minimal model are two particle collisions
between one electron or ion and one neutral particle, therefore
the intrinsic parameters simply scale with neutral particle density $N$.
Applying the minimal model to a different gas type at different temperature
or density can simply be taken into account by adapting
the intrinsic scales (\ref{Sc1})--(\ref{Sc2}), while the dimensionless
model (\ref{nodim})--(\ref{poissnd}) stays unchanged. 
This is true for the fast two particle processes in the front,
for corrections compare, e.g., \cite{LiuPasko,LiuPasko2}.

\subsection{Additional mechanisms: photo
and background ionization, gas convection}

The present model is suitable to describe negative or anode
directed streamers in gases with negligible photo-ionization 
like pure nitrogen.
Negative streamers move in the same direction as the electrons drift.
It should be noted that they hardly have been
studied experimentally in recent years except by Yi and
Williams \cite{YiWilliams} who carefully studied the influence of
small oxygen concentrations on streamers in nitrogen. 
Their negative streamers approach 
an oxygen-independent limit when the oxygen concentration becomes
sufficiently small. This supports our view that the minimal model
suffices to describe negative streamers in pure nitrogen.
 
However, positive streamers in nitrogen do depend even on
very small oxygen concentrations \cite{YiWilliams}. 
This can also be understood:  
The head of a positive or cathode directed streamer would be 
depleted from electrons within the minimal model and hardly move
while a negative streamer keeps propagating through electron drift
\cite{PRLUWC,PREUWC}.
On the other hand, photo-ionization or a substantial amount 
of background ionization supply electrons ahead of the streamer tip
and do allow a positive streamer to move. A recent discussion of
the relative importance of these two mechanisms has been given 
in \cite{Pan}. The main conclusion is that in repetitive discharges,
the background ionization is important.

A very interesting feature contained in none of these models
is gas convection. Typically, it is assumed that the relative 
density of charged particles $n_e/N$ is so small --- according to
dimensional analysis (\ref{Sc2}), it is of the order of $10^{-5}$
under normal conditions --- that the neutral gas stays at rest.
However, streamer induced gas convection recently has become 
a relevant issue in airplane hydrodynamics! Actually,
streamers might efficiently accelerate the air in the boundary layer 
above a wing and therefore decrease the velocity gradient and
allow the flow over the wing
to stay laminar. For these exciting results with relevance
also for general streamer studies, we refer to \cite{PitchW,StariW}.
We finally remark that this effect should not be relevant
for sprite discharges, as the characteristic degree of ionization
is $2\cdot10^{-5}\;N/N_0$ according to dimensional analysis,
so it decreases to the order of $10^{-10}$ at 70 km height.


\section{Numerical solutions of the minimal model}

\subsection{The stages of streamer evolution}

\label{subsecstages}
\begin{figure}
\begin{flushright}
\includegraphics[width=13cm]{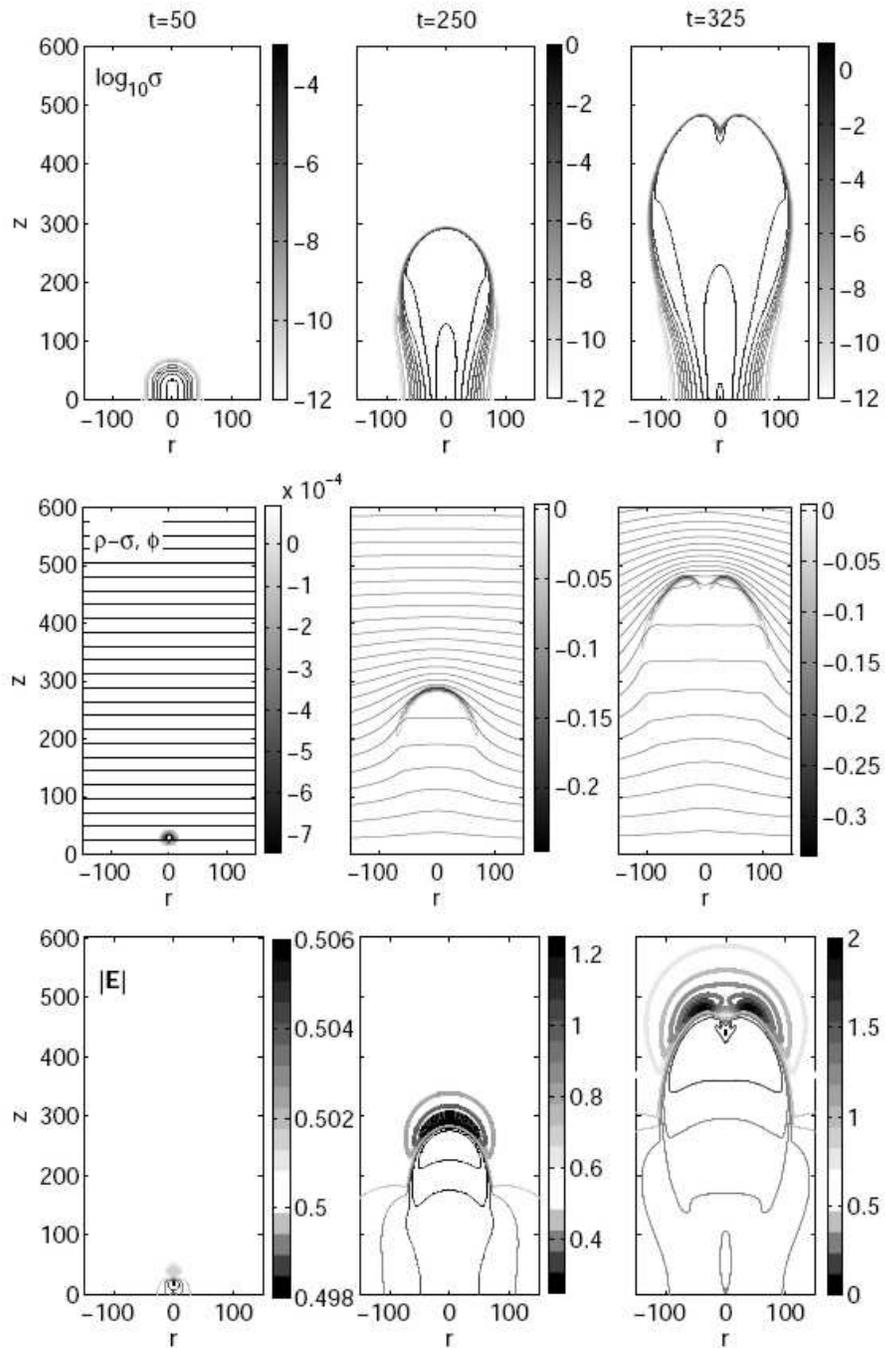}
\caption{Evolution of the electron density distribution (upper row), 
net charge density and equipotential lines (middle row) 
and electric field strength (lower row), when
an ionization seed attached to the cathode ($z$=0) is released in a 
background field of 0.5 in dimensionles units. 
The thick lines in the lower panel indicate
where the field is higher than the background value, 
the thin lines where it is lower. The
snapshots are taken at dimensionless times $t$=50, 250 and 325.
{\it Note that the quality and size of this figure (and some others) 
has been substantially reduced for the arxiv submission and 
will be better in the journal publication.}}
\label{streamerevol}
\end{flushright}
\end{figure}

The solutions of the minimal model in Fig.~\ref{streamerevol} 
show the characteristic
states of streamer evolution. Shown is a negative streamer in a high background
field of 0.5 in dimensionless 
units as presented in \cite{PRLMan,PREStrAndr,ICPIGCaro,CaroThesis,JCP}.

In the first column, a streamer just emerges
from an avalanche. At this stage the space charges are smeared out
over the complete streamer head and
resemble very much the historical sketches of Raether \cite{Raether}
(cf.~Fig.~\ref{Raether}(a) below), 
as they also can be found in the textbooks of Loeb and Meek
\cite{LoebMeek} or Raizer \cite{RaizerGas}.

In the second column, the space charge region has contracted to a thin layer
around the head, as also can be seen in many other studies of positive
or negative streamers in nitrogen or air; now the electric field
is suppressed in the ionized
interior and substantially enhanced ahead of the streamer tip.
As Raether's estimate in \cite{Raether} shows (cf.~Fig.~\ref{Raether}(a)), 
the field cannot
be substantially enhanced in the initial stage of streamer evolution.
However, it can in the second stage when the thin layer is formed.
In this stage, the enhanced field in our calculation also easily can exceed
the theoretical value suggested by Dyakonov and Kachorovskii
\cite{DK1,DK2}, and by Raizer and Simakov \cite{RaizerSim}.

In the third column, the streamer head becomes unstable and branches.
When we first published these results in \cite{PRLMan}, doubts
were raised about the physical nature of the branching event
\cite{PRLKuli,PRLReply}, refering also to an earlier debate
of a similar observation \cite{Pan2003,KuliBr,Pan2001}. 
This debate motivates our analysis and discussion in Section 5.

\subsection{The multiscale nature of streamers}

\begin{figure}
\begin{center}
\includegraphics[height=10cm]{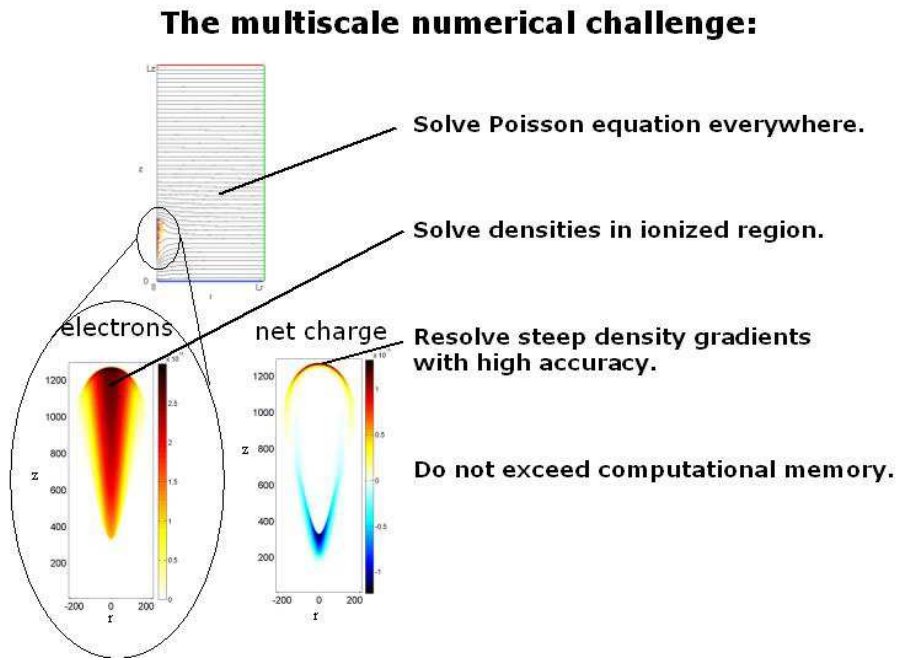}
\caption{Sketch of the inherent scales of a propagating streamer.}
\label{multiscale}
\end{center}
\end{figure}

It is important to note the very different inherent scales of a
propagating streamer, even within the minimal model. They are shown
in Fig.~\ref{multiscale}: there is a wide non-ionized outer space 
where only the
electrostatic Laplace equation $\nabla^2\phi=0$ has to be solved
to determine the electric field. There are one or several streamer
channels that are long and narrow. Around the streamer head,
there is a layered structure with an ionization region and a
screening space charge region.

Furthermore, in future work more regions should be distinguished:
there is the leading edge of the front where the particle density
is so low that the stochastic particle distribution leads to substantial
fluctuations, and there is the interior ionized region where
statistical fluctuations of particle densities are negligible:
the characteristic number of charged particles within a characeristic 
volume $\alpha_0^{-3}$ is according to dimensional analysis
\be
\frac{\epsilon_0\alpha_0{\cal E}_0}{\rm e}\cdot\alpha_0^{-3}
\approx\frac{6 000}{N/N_0}.
\ee
An immediate consequence is that stochastic density fluctuations 
are more important for high pressure discharges like streamers 
than for sprites. Whether this leads to different branching rates
has to be investigated.

A large separation of length scales can be a benefit for analysis
as it allows to use their ratios as small parameters and to
develop a ladder of reduced models, see Section 5. On the other hand,
it is a major challenge for numerical calculatioms.

\subsection{Adaptive grid refinement}

\begin{figure}
\begin{center}
\includegraphics[width=12cm]{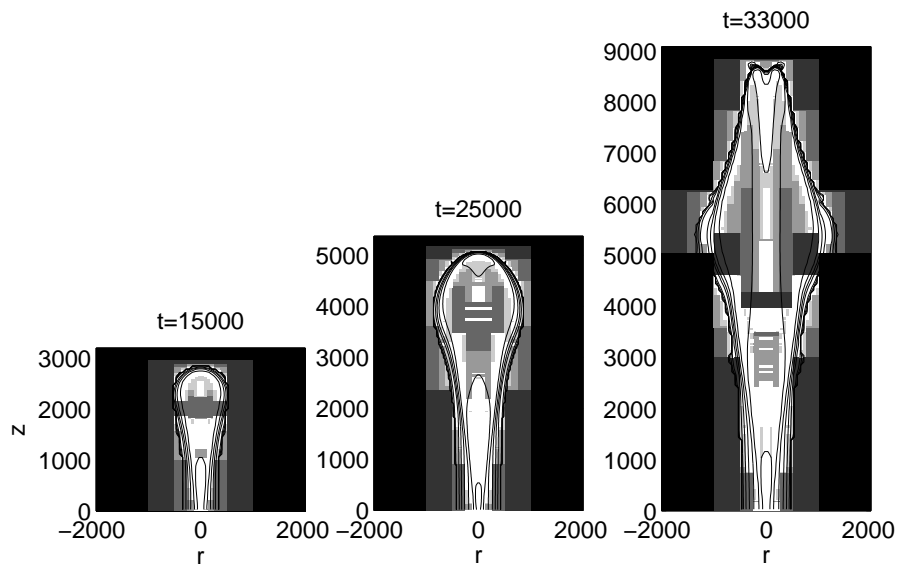}
\caption{Evolution of the logarithm of the electron density together with 
the computational grids for a streamer evolving in a background field
of 0.15, with an inter-electrode distance of approx. 65000.
In pure molecular nitrogen under normal conditions this corresponds to
a $\approx$15 cm gap with a background field of 30 kV/cm. The initial seed is attached 
to the cathode, allowing for a net electron inflow. The coarsest grid, in black, has a mesh size
of 64, and is refined up to a mesh size of 2 (white domains).  }
\label{grids}
\end{center}
\end{figure}

These numerical challenges can be met by adaptive grid refinement.
Such a code has recently been constructed. 
It computes the evolution of a streamer 
on a relatively coarse grid and refines the mesh where the fine 
spatial structure of the solution requires. 
For preliminary and more extended results,
we refer to \cite{IEEECaro,ICPIGCaro,CaroThesis,JCP} and future papers.
The distribution of the grid at different time steps is illustrated in
Fig.~\ref{grids}. Here the evolution of a streamer in a long, undervolted gap
is shown. More specifically, we consider a plane parallel electrode geometry,
with an inter electrode distance of approximately 65000 in dimensionless units.
The applied background electric field is uniform, and has a strength of 0.15.
For N$_2$ under normal conditions this corresponds to a gap of about 15 cm 
with a background field of 30 kV/cm.

We remark that the grid size used in the computation of the results shown
in Fig.~\ref{grids} is the same as that used by Vitello {\em et al.}~\cite{Vitello}
for a much smaller gap (0.5 cm) at a higher background field (50 kV/cm). In these
simulations streamer branching was not seen, probably due to
the shortness of the gap. On the contrary, a large gap as in the above examples
enables the streamer to reach the instability even at a relatively low field. 

Moreover, our previous calculations that showed streamer
branching in a high background field of 0.5, were performed on a 
uniform numerical grid with grid spacing $\Delta x=2$
in \cite{PRLMan} and with $\Delta x=1$ in \cite{PREStrAndr}.
We now are able to perform computations on a grid that adapts
locally down to $\Delta x=1/8$ \cite{CaroThesis,JCP}.

For both background fields 0.5 and 0.15, the numerical results show
that the time of streamer branching reaches a fixed value 
when using finer numerical meshes. 
We therefore conclude that streamer branching indeed is physical,
and we will further support this statement with different arguments 
in the next section. We notice that in our cylindrically symmetric system, 
the streamer branches into rings,
which obviously is rather unphysical. Therefore it is not meaningful to follow
the further evolution of the streamer after branching. However, 
the effectively two-dimensional setting suppresses destablizing modes 
that break the cylindrical symmetry,
and the time of branching in a cylindrically symmetric system 
therefore gives an upper bound for the branching time in the 
real three-dimensional system~\cite{PRLReply}.


\section{Analytical results on propagating and branching streamers}

\subsection{Nonlinear analysis of ionization fronts}

The question of streamer branching can be addressed analytically,
using concepts developed in other branches of science:
Combustion, e.g., for decades is a very active area of applied nonlinear 
analysis, pattern formation and large scale computations.
Chemical species are processed/burned when fuel is available
and the temperature exceeds a threshold. The temperature is
enhanced by the combustion front itself.
Quite similarly, ionization is created is there are free
electrons and if the electric field exceeds a threshold.
The field is enhanced by the (curved) ionization front itself
\cite{PRLUWC,PREUWC,PRLMan}.

It is therefore attractive to develop analysis of streamers along
the same lines, hence complementing numerical results with
an analytical counterpart. This is particularly important
for addressing questions of branching, long time evolution
and multi-streamer structures. In particular,
the structures of many interacting streamers in the near future 
will remain numerically inaccessible without model reduction.

\subsection{The moving ionization boundary}

In recent work, we have elaborated streamer evolution on 
two levels of refinement: the properties of a planar ionization
front within the minimal model \cite{PRLUWC,PREUWC,PRLMan,Man04},
and the evolution of curved ionization boundaries \cite{IEEECaro,Bern1,Bern2}.
A front solution is a solution of the full fluid model 
(\ref{nodim})--(\ref{poissnd}) zooming into the inner structure
of the front. Ionization boundaries are formulated on the outer scale
where the ionization front is reduced to a moving boundary 
between ionized and nonionized region.

If there is no initial ionization in the system,
planar negative ionization fronts within the minimal model 
move with asymptotic velocity
\be
v^*(E^+)=|E^+|+2\sqrt{D\;f(|E^+|,\nu)}
\ee
into a field $E^+$ immediately ahead of the front. 
The degree of ionization $\sigma^-=\rho^-$ behind the front
is a function of the field $E^+$. For large fields,
the front velocity is dominated by the electron drift velocity
$v^*(E^+)\approx|E^+|$. For details about analyzing streamer fronts, 
we refer to \cite{PRLUWC,PREUWC,PRLMan,Man04}. 

\begin{figure}
\begin{center}
\includegraphics[width=10cm]{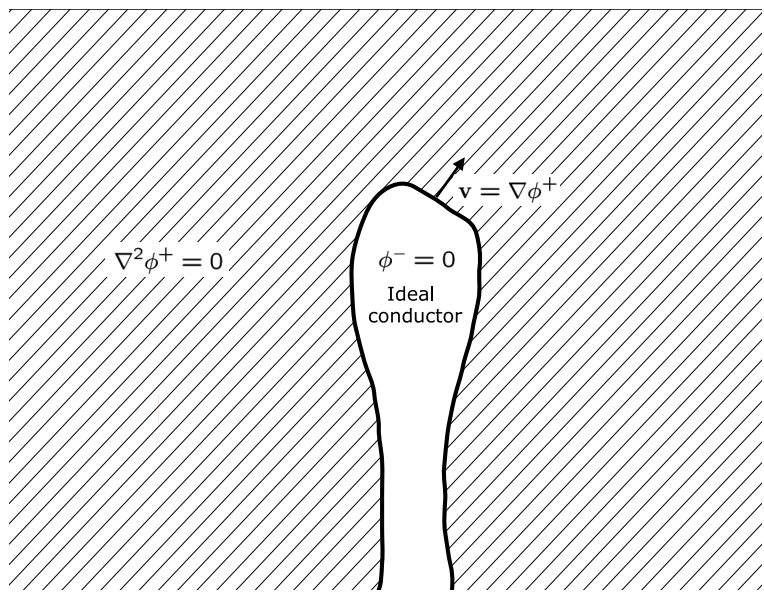}
\caption{The moving boundary approximation for an ideally conducting
streamer. }
\label{MBA}
\end{center}
\end{figure}

On the outer level of ionized and non-ionized region, a simple evolution
model for the phase boundary can be formulated as shown in Fig.~\ref{MBA}: 
assume the Lozansky
Firsov approximation \cite{Firsov} that the streamer interior (indicated
with an upper index $^-$) is equipotential: $\phi^-=$ const.
The exterior is free of space charges, hence $\nabla^2\phi^+=0$.
Every piece of the boundary moves with the local velocity $v^*(E^+)$
determined by the local field $E^+$ ahead of the front.

\begin{figure}
\begin{center}
\includegraphics[height=4.5cm]{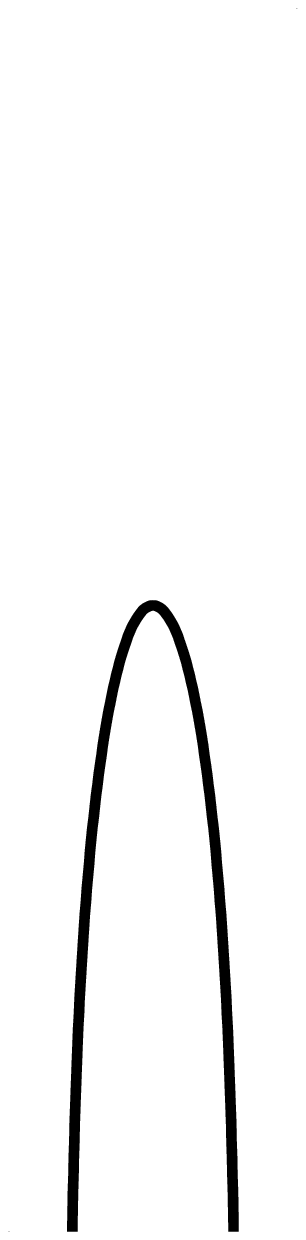}
\includegraphics[height=4.5cm]{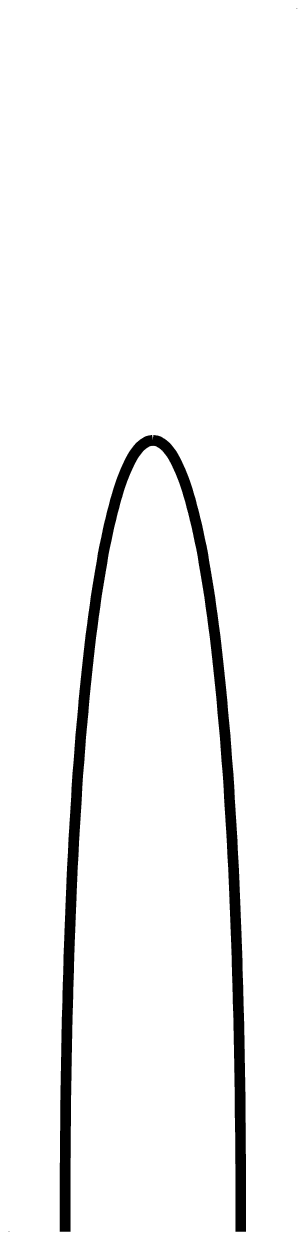}
\includegraphics[height=4.5cm]{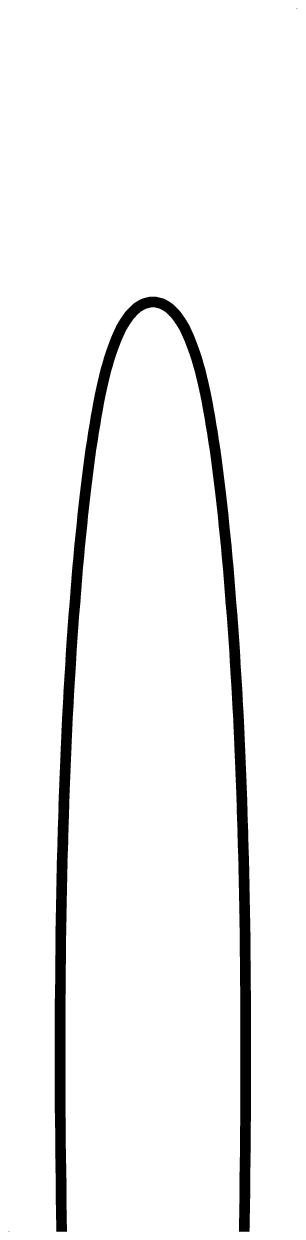}
\includegraphics[height=4.5cm]{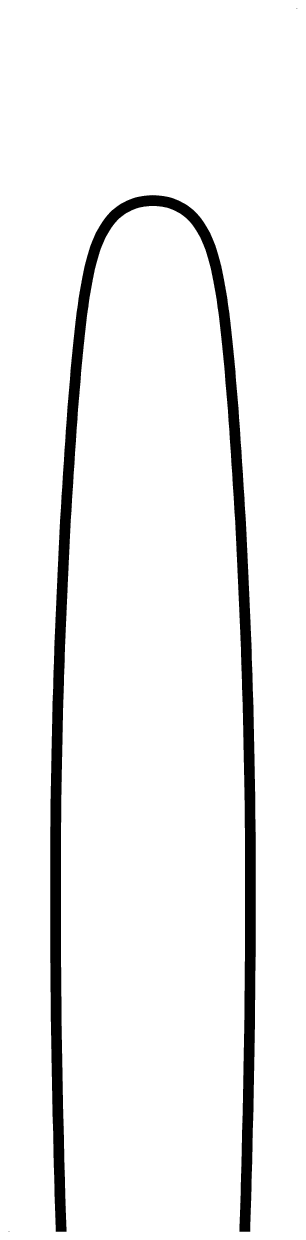}
\includegraphics[height=4.5cm]{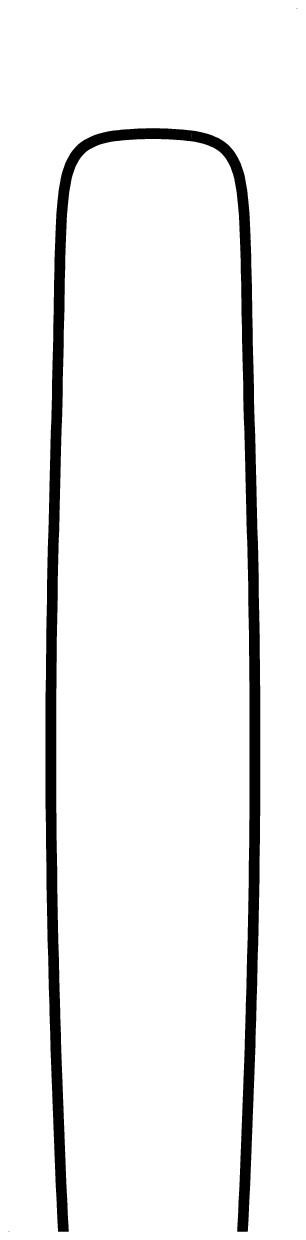}
\includegraphics[height=4.5cm]{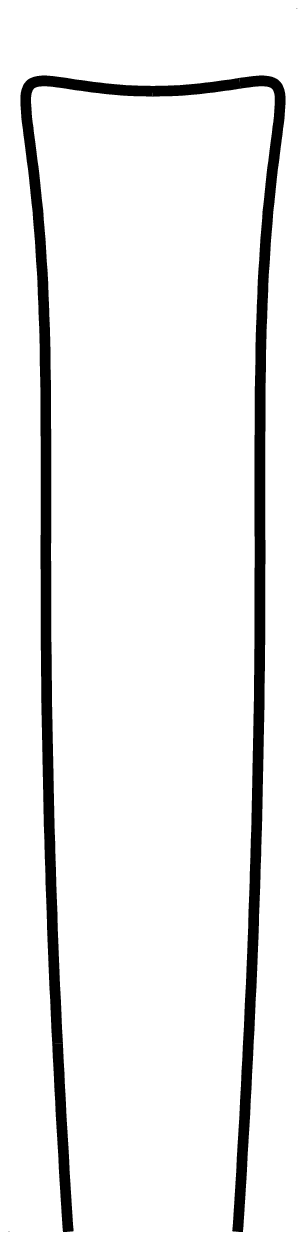}
\caption{Temporal evolution of the streamer in the moving boundary 
approximation 
for an ideally conducting body, as shown in Fig.~\ref{MBA}. The
solution was computed assuming that the electric potential accross 
the ionization boundary is continuous $\phi^-=\phi^+$ and 
using conformal mapping methods~\cite{Bern1}.}
\label{Bernbr}
\end{center}
\end{figure}

If one assumes that the electric potential across the ionization
boundary is continuous $\phi^+=\phi^-$ everywhere along the boundary,
one arrives at a model that has been studied previously in 
the hydrodynamic context of viscous fingering. Our solutions 
of the model \cite{IEEECaro,Bern1} show that the transition 
from convex to concave streamer head indeed is dynamically possible
(see Fig.~\ref{Bernbr}); this is the onset of 
streamer branching. While these solutions demonstrate the
onset of branching, they have the unphysical property that the local
curvature of the boundary can become infinite within finite time.
This cusp formation is suppressed in viscous fingering by a regularizing
boundary condition. Our analysis \cite{Bern2}
of streamer fronts suggests a new boundary condition
\be
\phi^+-\phi^-=F(|E^+|),~~~F(|E^+|)\approx |E^+| ~~~{\rm for}~|E^+|>2,
\ee
for the potential jump accross the boundary. This boundary condition
aproximates the pde's (\ref{sigmand})--(\ref{poissnd}); it can be 
understood as a floating potential on the non-ionized side
of the ionization boundary, if the potential on the ionized side is fixed.
First results \cite{Bern2} with purely analytical methods indicate 
that this boundary condition indeed prevents cusp formation, i.e.,
it regularizes the problem.

We conclude that streamer branching is generic even for deterministic
streamer models when they approach a state when the width of the
space charge layer is much smaller than its radius of curvature,
as in the second and third column of Fig.~5. A sketch
of the distribution of surface charges and field is given by Fig.~8.
A streamer in this state is likely to branch due to 
a Laplacian instability. A more precise characterization of the
unstable state of the streamer head is under way. 

\subsection{How branching works and how it doesn't}

\begin{figure}
\begin{center}
\begin{tabular}{cc}
$\begin{array}{c}
\includegraphics[height=4.5cm]{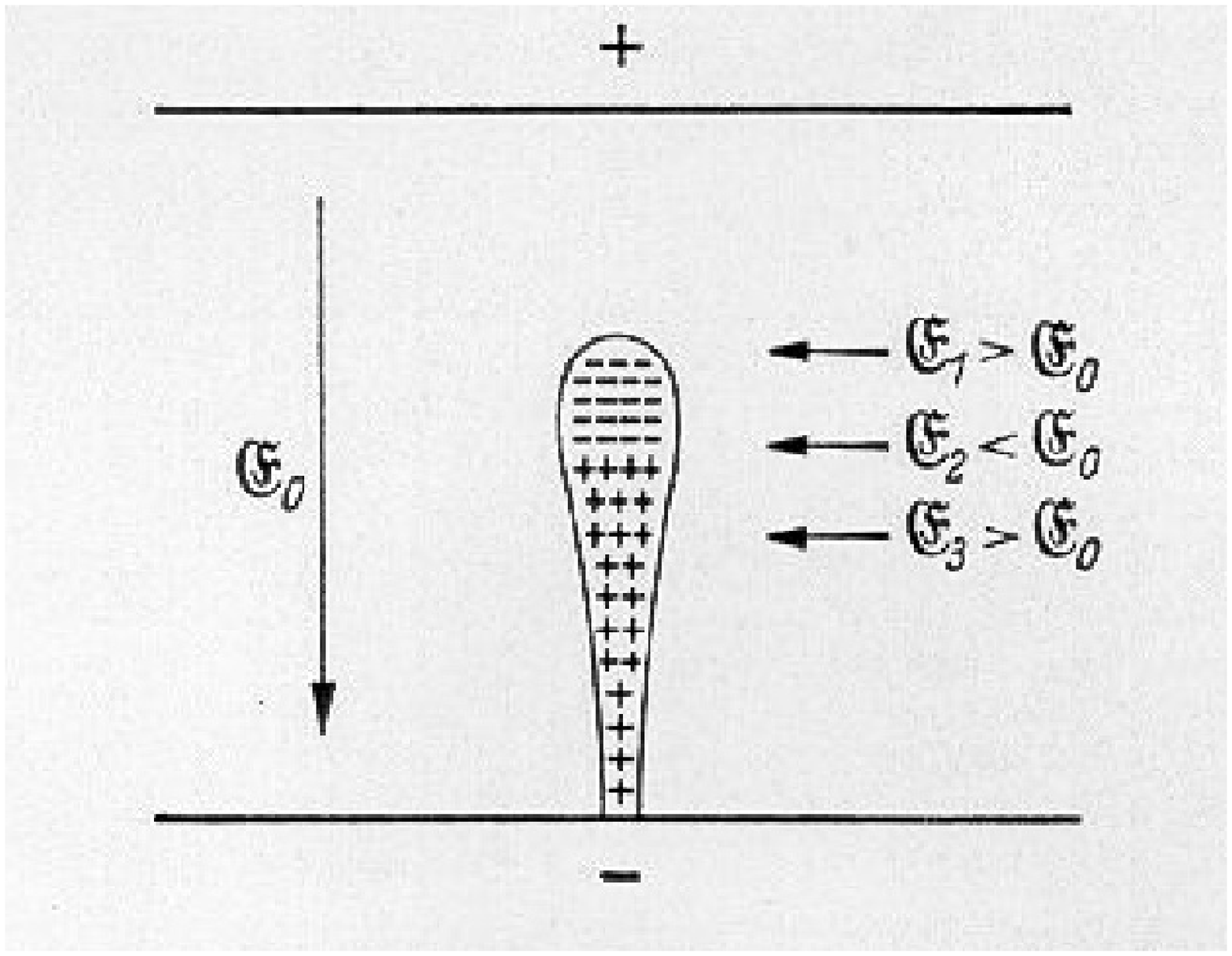}\\
{\rm (a)}\\
\includegraphics[height=3.5cm]{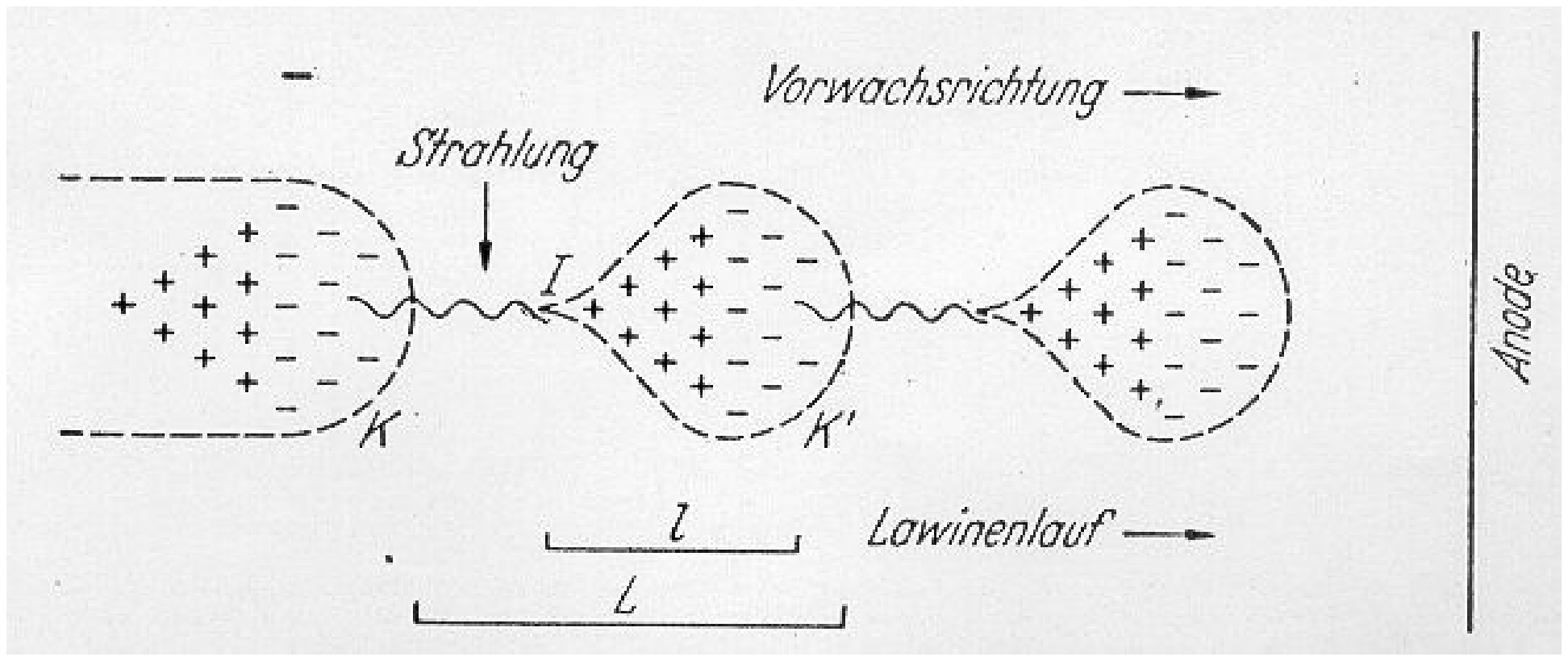}\\
{\rm (b)}
\end{array}$ &
$\begin{array}{c}
\includegraphics[height=5.5cm]{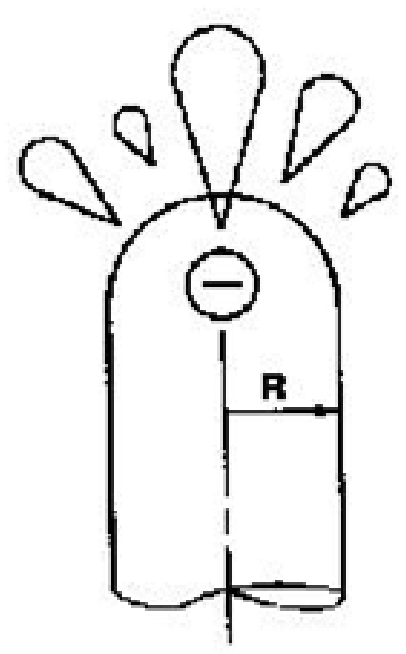}\\
{\rm (c)}
\end{array}$
\end{tabular}
\caption{(a) Space charge distribution in the emerging streamer head
and (b) streamer propagation by photoionization according to Raether 1939
\cite{Raether}. Fig.\ (b) was reproduced identically in \cite{LoebMeek}
with English labels. Figure (c) shows the pictorial concept
of streamer branching by rare and long-range photoionization
events and successive avalanches as it can be found in many textbooks
like \cite{Raizer}.
It is a minor variation of Fig.\ b with the avalanches not aligned, 
but placed around the streamer head. The present version of the figure
is taken from \cite{DBM2} where it motivates the dielectric breakdown
model \cite{DBM1}. Find our critique of this concept in the text.}
\label{Raether}
\end{center}
\end{figure}

It came as a surprise to many that a fully deterministic model like our
fluid model would exhibit branching, since another branching concept 
based on old pictures of Raether \cite{Raether} is very well known.
It is illustrated in Fig.~\ref{Raether}. We remark on this concept:\\
1) The distribution and shape of avalanches ahead of the streamer
head as shown in Fig.~\ref{Raether}~c to the best of our knowledge 
have never been substantiated by further analysis. \\
2) Even if this avalanche distribution is
realized, it has not been shown that it would evolve into
several new streamer branches. Our major point of critique is 
that a space charge distributed over the full streamer head as 
in the figure would be self-stabilizing and not destabilizing, 
cf.~a comparable recent analysis \cite{Jom}.

On the contrary, our main statement is:
{\em The formation of a thin space charge layer is necessary 
for streamer branching while stochastic fluctuations are not necessary.}

Furthermore, for the question whether branching is possible in a deterministic 
fluid model, we remind the reader that chaos is possible in fully 
deterministic nonlinear models when their evolution approaches
bifurcation points. Moreover, tip splitting in Laplacian 
growth problems as described in Fig.\ 8 is well established 
in viscous fingering in two fluid flow and other branches of physics.

\subsection{The importance of charge transport 
and a remark on dielectric breakdown models}

We have identified a state of the streamer head where it
can branch. This state is characterized by a weakly curved
ionization front, i.e., the radius of curvature is much larger
than the width of the front. The width of the front is
determined by the field $E^+$ ahead of it \cite{PREUWC}, 
for high fields the width saturates \cite{PREUWC,Man04}.
The formation of such a weakly curved front requires 
a sufficiently high potential difference between streamer tip 
and distant electrode and an appropriate charge content of the
streamer head. Charge is a conserved quantity. The consideration 
of charge conservation misses in the streamer concepts suggested 
in \cite{DK1,DK2,RaizerSim}.

These considerations lead us to the general idea that a 
streamer tip is characterized by electric potential $\phi$, 
curvature $R$, field enhancement $E^+$ and total charge content $Q$.
Only two of these four parameters are independent.

In contrast, dielectric barrier models (DBM) \cite{DBM1,DBM2}  
for multiply branched discharge structures are characterized
only by potential and longitudinal spatial structure.
We argue that a quantitative DBM model should also include 
the width of the streamer channel and the related charge content 
as a model variable. This would allow an appropriate 
characterization of streamer velocity and branching probability.

We remark that studies of streamer width and charge content 
are also crucial for determining the electrostatic interaction
of streamers --- or of leaders as their large relatives.

\section{Summary and outlook}

The purpose of the present paper was to review the presently available
methods to investigate streamer discharges, in particular, those
methods presently developed at TU Eindhoven and CWI Amsterdam.
The paper summarizes a talk given at the XXVII'th International 
Conference on Phenomena in Ionized Gases (ICPIG) 2005.

Obviously, the complexity and the many scales of the phenomenon
pose challenges to experiments, simulations, modeling and 
analytical theory if one wants to proceed to the quantitative
understanding of more than a single non-branching streamer. 
In the present stage, we have
developped reliable methods in each discipline, they are reviewed in
the present paper. In the next stage, results of different methods
should be compared: simulations should be compared with experiments,
and simulations should be checked on consistency with analytical
results. Analysis can also be used to extrapolate tediously generated
numerical results, once the emergence of larger scale coherent
structures --- like complete streamer heads with their inner layers
--- has been demonstrated. 

We have reviewed nanosecond resolved measurements of streamers and
the surprising influence of the power supply, streamer applications 
and the relation to sprite discharges above thunderclouds.
We have summarized the physical mechanisms of streamer formation
and a minimal continuum density model that contains the essentials 
of the process. We have shown that a propagating and branching
streamer even within the minimal continuum model consists of
very different length scales that can be appropriately simulated
with a newly developed numerical code with adaptive grids. Finally, 
we have summarized our present understanding of streamer branching 
as a Laplacian instability and compared it to earlier branching
concepts.

On the experimental side, future tasks are precise measurements 
of streamer widths, velocities and branching characteristics 
and their dependence on gas type and power supply as well
as quantitative comparison of streamers and sprites.
On the theoretical side, both microscopic and macroscopic
models should be developed further. Microscopically, the particle
dynamics in the limited region of the ionization front will 
be investigated in more detail. Macroscopically, the quantitative
understanding of streamer head dynamics should be incorporated
into new dielectric breakdown models with predictive power.
In particular, charge transport and conservation should be included.
The theoretical tasks can only be treated succesfully,
if a hierarchy of models on different length scales is developed:
from the particle dynamics in the streamer ionization front up
to the dynamics of a streamer head as whole. Elements of such
a hierarchy are presented in the present paper. 
\\

{\bf Acknowledgements:}
This paper summarizes work by a number of researchers in a number
of disciplines, therefore it has a number of authors. Beyond that,
we acknowledge inspiration and thought exchange with the pattern
formation group of Wim van Saarloos at Leiden Univ., with 
numerical mathematicians at cluster MAS at CWI Amsterdam, 
with colleagues in physics and electroengineering at TU Eindhoven, 
as well as with the many international colleagues and friends
whom we met at conferences on gas discharges, atmospheric discharges 
and nonlinear dynamics in physics and applied mathematics. 
We thank Hans Stenbaek-Nielsen and Elisabeth Gerken for making 
sprite figures 3 and 4 available.

The experimental Ph.D.\ work of Tanja Briels in Eindhoven is supported
by a Dutch technology grant (NWO-STW), the computational Ph.D.\ work
of Carolynne Montijn is supported by the Computational Science program
of FOM and GBE within NWO. The analytical Ph.D.\ work of Bernard
Meulenbroek was supported by CWI. The postdoc position of Andrea Rocco
was payed by FOM-projectruimte and the Dutch research school 
``Center for Plasma Physics and Radiation Technology'' (CPS).

\end{document}